\documentclass[journal]{IEEEtran}

\usepackage[cmex10]{amsmath}
\usepackage{amssymb}
\usepackage{amsfonts}
\usepackage{bbm}

\def\beq{\begin{equation}}
\def\eeq{\end{equation}}
\def\be{\begin{equation}}
\def\ee{\end{equation}}
\def\ben{\begin{eqnarray}}
\def\een{\end{eqnarray}}

\newcommand{\Tr}{\operatorname{Tr}}
\newcommand{\bra}[1]{\langle #1 |}
\newcommand{\ket}[1]{| #1 \rangle}

\newcommand{\pro}[1]{\ket{#1}\bra{#1}}

\def\openone{\mathbb{I}}

\def\tr{{\rm Tr}}
\def\id{{\rm id}}

\def\>{\rangle}
\def\<{\langle}

\def\ot{\otimes}

\newtheorem{lemma}{Lemma}

\begin{document}
%
% paper title
% can use linebreaks \\ within to get better formatting as desired
\title{On quantum advantage in dense coding}

%
%
% author names and IEEE memberships
% note positions of commas and nonbreaking spaces ( ~ ) LaTeX will not break
% a structure at a ~ so this keeps an author's name from being broken across
% two lines.
% use \thanks{} to gain access to the first footnote area
% a separate \thanks must be used for each paragraph as LaTeX2e's \thanks
% was not built to handle multiple paragraphs
%

\author{Micha{\l}~Horodecki~and~Marco~Piani%
\thanks{M. Horodecki is with the Institute of Theoretical Physics and Astrophysics, University of Gda{\'n}sk, 80--952 Gda{\'n}sk, Poland}%
\thanks{M. Piani was with the Institute of Theoretical Physics and Astrophysics,
University of Gda{\'n}sk, 80--952 Gda\'nsk, Poland when this work was done, and is now with the
Institute for Quantum Computing \& Department of Physics and Astronomy,
University of Waterloo, 200 University Ave. W., N2L 3G1 Waterloo, Canada}}%

\maketitle

\begin{abstract}
%\boldmath
The quantum advantage of dense coding is studied, considering general encoding quantum operations. Particular attention is devoted to the case of many senders, and it is shown that restrictions on the possible operations  on the senders' side may make some quantum state useless for dense-coding. It is shown, e.g., that some states are useful for dense coding if the senders can communicate classically (but not quantumly), yet they cannot be used for dense coding, if classical communication is not allowed. These no-go results are actually independent of the particular quantification of the quantum advantage, being valid for any reasonable choice. It is further shown that the quantum advantage of dense coding satisfies a monogamy relation with the so-called entanglement of purification.
\end{abstract}
% IEEEtran.cls defaults to using nonbold math in the Abstract.
% This preserves the distinction between vectors and scalars. However,
% if the journal you are submitting to favors bold math in the abstract,
% then you can use LaTeX's standard command \boldmath at the very start
% of the abstract to achieve this. Many IEEE journals frown on math
% in the abstract anyway.

% Note that keywords are not normally used for peerreview papers.

% For peer review papers, you can put extra information on the cover
% page as needed:
% \ifCLASSOPTIONpeerreview
% \begin{center} \bfseries EDICS Category: 3-BBND \end{center}
% \fi
%
% For peerreview papers, this IEEEtran command inserts a page break and
% creates the second title. It will be ignored for other modes.
%\IEEEpeerreviewmaketitle

\section{Introduction}
\label{sec:introduction}

Entanglement~\cite{horodecki2009quantum} plays a central role in quantum information theory~\cite{NC,Alber2001}, especially in quantum communication. It is a physical resource exploited in tasks like teleportation~\cite{teleportation} and dense coding~\cite{BennettWiesner}. In the last communication problem, the sender, Alice, and the receiver, Bob, share a pair of two-level systems, or qubits, in a maximally entangled state
\beq
\label{eq:maxent2qubits}
\ket{\psi_0}=\frac{\ket{00}+\ket{11}}{\sqrt{2}}.
\eeq
Alice can transmit two classical bits of information sending her qubit to Bob, i.e. with a exchange of just one qubit. To achieve this result, Alice sends her qubit after having applied an appropriate unitary rotation, corresponding for example to the identity $\sigma_0$ and the Pauli matrices $\sigma_i$, $i=1,2,3$.
The resulting states $\ket{\psi_\mu}=(\sigma_\mu\otimes 1)\ket{\psi_0}$ (Bell states) are orthogonal, so that when Bob has received Alice's subsystem they can be unambiguously distinguished.

The previous result is possible because the two parties share initially an entangled state. Indeed, the Holevo bound~\cite{holevo} implies that one qubit may carry at most one classical bit of information, if no pre-established (quantum) correlations between the parties exist.

Unfortunately, in real world applications we have to deal with imperfect knowledge and noisy operations, therefore the resulting (shared) quantum states are mixed and described in terms of density matrices. From the point of view of quantum information, this is most relevant in the distant-labs paradigm. In this case two (or more) parties may want to share a maximally entangled system, but their actions are limited to local operations and classical communication (LOCC), so that they cannot create shared entangled states, but only process by LOCC pre-existing (imperfect) quantum correlated resources. One possible way out is entanglement distillation~\cite{BBPS1996}, which can be realized by LOCC. On the other hand, the task in dense coding is exactly that of communicating classical information in the most efficient way with the exchange of a quantum system, hence we should not allow classical communication between the parties~\footnote{Another possibility is that of \emph{counting} communication, considering consumed and total communication, and computing the achieved communication as the difference, but we will not consider such a framework.}.

Therefore, it is interesting to study coding protocols for Alice to send classical information to Bob, directly acting on copies of a shared mixed state $\rho_{AB}$. The problem was further generalized in ~\cite{bruss, bruss2}, where the notion of distributed quantum dense coding was introduced: in that picture, many senders, called Alices, share states with many receivers, called Bobs. In ~\cite{bruss, bruss2} the encoding is purely unitary, as in the standard pure two-parties setting, i.e. a letter in an alphabet is associated to a unitary operation.
With this protocol, it was shown that,  for a given network scheme---i.e., for a given choice of which Alice sends her subsystem to which Bob---the possibility of dense coding does not depend on the allowed operations among the senders, but does depend on the allowed operations among the receivers, i.e. on the allowed decoding processes. In the case global operations are allowed among the Bob's, i.e. there is essentially only a single receiver, dense coding is possible if and only if the coherent information~\cite{SchumNiel} between the Alices and the Bobs,
\beq
I(A\>B)=S(\rho_B) - S(\rho_{AB}),
\eeq
with $S(\sigma)=-\tr\rho \log \rho$ the von Neumann entropy, is strictly positive. Both $I(A\>B)$ and $I(B\>A)$ are less or equal to zero for separable or bound entangled states \cite{HH94-redun}. Indeed, if $I(A\>B)$ or $I(B\>A)$ is strictly positive, we know that the state $\rho_B$ is distillable thanks to the \emph{hashing inequality}~\cite{hashinginequality}. Therefore neither separable nor bound entangled states (along the cut A:B) are useful for dense coding.

In the present paper we consider a
general encoding scheme: each letter in an alphabet is associated to a completely positive trace preserving (CPTP) map. Such a scheme was already presented in~\cite{michal,winterdense}, in the one-sender-one-receiver context, where it was found that the optimal encoding is still unitary, but only after a ``pre-processing'' operation which optimizes the coherent information $I(A\>B)$ between the parties and is independent from the letter of the alphabet to be sent. Indeed, while no CPTP operation by Bob can increase $I(A\>B)$~\footnote{Mathematically, this is immediately proved by means of strong subadditivity of von Neumann entropy. Moreover, it is consistent with the fact that in the decoding process the most general operations by Bob are allowed, so that any pre-processing operation by Bob can be thought to happen \emph{after} he has received the other half of the state from Alice.}, an operation by Alice can. The simplest example was given in~\cite{michal}: the sender discards a noisy subsystem of hers, which is factorized with respect to the remnant state and as such can not increase the capacity, i.e., the maximal rate of transmission.

We show that, as a consequence, in the case of multi-senders, the capacity depends on the allowed operations among the senders because this may restrict the pre-processing operation to be non-optimal (with respect to the global operation, i.e. to the one-sender setting). This may be understood considering the case where the noise to be traced out to increase the coherent information, is not concentrated in some factorized subsystem, but spread over many subsystems; it may be possible to concentrate and discard such noise with some global (on Alice's side) operation, but not with local ones.

In~\cite{bruss, bruss2} a classification of quantum states according to their usefulness for distributed dense coding with respect to the allowed operation on the receiver's side was depicted. Considering pre-processing, a similar classification can be made with respect to the allowed operations on the sender's side. We provide concrete examples of states which are useful for dense coding only if, e.g., global operations are allowed among the senders -- i.e., there is only one sender -- but not if they are limited to LOCC (similarly, there are states that are useful only if LOCC are allowed, but not if the senders are limited to local operations). We remark that the constraints on the usefulness of a given multipartite state, based on the allowed operations, persist even in the most general scenario of preprocessing on many copies.

As a general observation about dense coding, we further show that, in the basic one-sender-one-receiver scenario, what we call the quantum advantage of dense coding satisfies a monogamy relation with the so-called entanglement of purification.

The paper is organized as follows. In Section \ref{sec:two-party} we define the terms of the problem and present a formula for the capacity of two-party dense coding with pre-processing. In Section \ref{sec:examples} we move to the many-senders-one-receiver setting, specifying the various classes of operations that we will allow among the senders. In Section \ref{sec:exampleshierarchy} we analyze the many-senders-one-receiver  setting, giving examples of how the dense-codeability of multipartite states depends on the allowed pre-processing operations among the senders, particular providing sufficient conditions for non-dense-codeability with restricted operations. In Section \ref{sec:symmext} we discuss how dense-codeability is related to distillability and the concept of symmetric extensions. In Section \ref{sec:manycopies} we briefly consider the asymptotic setting, and argue that the limits to dense-codeability presented in the previous sections remain valid. In Section \ref{sec:monogamy} we elaborate on  the monogamy relation between what we call the quantum advantage of dense coding and a measure of correlations known as entanglement of purification. Finally, we discuss our results in Section \ref{sec:discussion}.

\section{Two-party dense coding with pre-processing}
\label{sec:two-party}

We rederive here some known results for the two-party case, i.e. we allow global operations on both sides (sender's and receiver's). Such results will be the backbone of our further discussion. Moreover, we discuss some subtle points of dense coding which arise already at the level of the two-party setting, and stress the importance of a quantity defined in terms of a maximization of coherent information, that we call \emph{quantum advantage of dense coding}.

Alice and Bob share a mixed state $\rho^{AB}$ in dimension $d_A\otimes d_B$, i.e. $\rho^{AB}\in \mathcal{M}(\mathbb{C}^{d_A})\otimes\mathcal{M}(\mathbb{C}^{d_B})$. The protocol we want to optimize is the following. (i) Alice performs a local CPTP map $\Lambda_i:\mathcal{M}(\mathbb{C}^{d_A})\rightarrow\mathcal{M}(\mathbb{C}^{d'_A})$ (note that the output dimension $d'_A$ may be different than the input one $d_A$) with \emph{a priori} probability $p_i$ on her part of $\rho^{AB}$. She therefore transforms $\rho^{AB}$ into the ensemble $\{(p_i,\rho^{AB}_i)\}$, with $\rho^{AB}_i=(\Lambda_i\otimes\id)[\rho^{AB}]$. (ii) Alice sends her part of the ensemble state to Bob.  (iii) Bob, having at disposal the ensemble $\{(p_i,\rho^{AB}_i)\}$, extracts the maximal possible information about the index $i$.

Notice that, for the moment being, we allow only one-copy actions, i.e. Alice acts partially only on one copy of $\rho^{AB}$ a time. On the other hand, we analyze the asymptotic regime where long sequences are sent. Moreover, notice that stated as it is, the protocol requires a perfect quantum channel of dimension equal to the output dimension $d'_A$, i.e. the ability to send perfectly a quantum system characterized by the mentioned dimension. We now prove that the capacity (rate of information transmission per shared state used) for this protocol corresponds to
\beq
\label{eq:capacity}
\begin{split}
C^{d_A'}(\rho^{AB})&=\log d'_A + S(\rho^B)-\min_{\Lambda_A}\big((\Lambda_A\otimes\id_B)[\rho^{AB}]\big)\\
				&=\log d'_A +\max_{\Lambda_A} I'(A\>B),
\end{split}
\eeq
where $I'(A\>B)$ is the coherent information of the transformed state $(\Lambda_A\otimes\id_B)[\rho^{AB}]$.
Note that the quantity depends both on the shared state $\rho^{AB}$ and on the output dimension $d'_A$ of the maps $\{\Lambda_i\}$, not only through the first logarithmic term, but also because the minimum runs over such maps.

We reproduce here essentially the same proof used in~\cite{michal}, but our attention will ultimately focus on the rate of communication \emph{per copy} of the state used, not on the capacity of a perfect quantum channel of a given dimension assisted by an unlimited amount of noisy entanglement. In this sense, our approach is strictly related to the one pursued in~\cite{winterdense}, as we will discuss in the following. We anticipate that there will be an important difference: in~\cite{winterdense} there is no optimization over the output dimension of $\Lambda_A$ (i.e. of the quantum channel), when it comes to consider the rate per copy of the state, with one use of the channel per copy.

The capacity (attained in the asymptotic limit where Alice sends long strings of states $\rho_i^{AB}$) of the dense coding protocol depicted above is given by the Holevo quantity
\beq
C^{d_A'}(\rho^{AB})=\max_{\{(p_i,\Lambda_i)\}}\Big(S(\sum_i\rho^{AB}_i)-\sum_ip_iS(\rho^{AB}_i)\Big).
\eeq

We bound it from above considering an optimal set $\{(\hat{p}_i,\hat{\Lambda}_i)\}$. Since the entropy is subadditive and no operation by Alice can change the reduce state $\rho^B$, we have
\beq
\begin{split}
C^{d_A'}(\rho^{AB})&\leq S(\sum_i\hat{\rho}^{A}_i) + S(\rho^B) - \sum_i p_i S(\hat{\rho}_i^{AB})\\
					&\leq \log d'_A+S(\rho^B)-\min_{\Lambda_A}S\big((\Lambda_A\otimes \id_B)[\rho_{AB}]\big).
\end{split}
\eeq
The quantity in the last line of the previous inequality, corresponding to \eqref{eq:capacity}, can be actually achieved by an encoding with $p_i=1/{d'_A}^2$ and $\Lambda_i[X]=U_i \Lambda_A[X] U_i^\dagger$. The unitaries $\{U_i\}_{i=1}^{{d'_A}^2}$ are orthogonal, $\tr(U_i^\dagger U_j)=d'_A\delta_{ij}$, and satisfy $1/d'_A \sum_i U_i X U_i^\dagger=\tr(X)\openone$, for all $X\in\mathcal{M}(\mathbb{C}^{d'_A})$, while the CPTP map $\Lambda_A$ corresponds to the pre-processing operation. Indeed, in this case $\sum_i\rho^{AB}_i=\openone/d'_A\otimes \rho^B$, so that
$S(\sum_i\rho^{AB}_i)=\log d'_A + S(\rho_B)$, and $\sum_ip_iS(\rho^{AB}_i)=S\big((\Lambda_A\otimes \id_B)[\rho_{AB}]\big)$. We notice in particular that Alice may always choose to substitute her part of the shared state with a fresh ancilla in a pure state. This corresponds to a pre-processing $\Lambda_{\rm sub}[X]=\tr(X)\pro{\psi}$ and gives a rate $\log d'_A$, corresponding to classical transmission of information with a $d'_A$-long alphabet, i.e. without a quantum advantage. A quantum effect is present if $\chi>\log d'_A$, i.e. if a local operation on Alice side is able to reduce the entropy of the global state strictly below the local entropy of Bob, or if $I(A\>B)>0$ from the very beginning.

An almost trivial case where preprocessing has an important role is $\rho^{AA'B}=\rho^{AB}\otimes\rho^{A'}$, with $S(\rho^{AB})<S(\rho^{B})$ but $S(\rho^{AB})+S(\rho^{A'})\geq S(\rho^{B})$. Here we consider $AA'$ as a composite system Alice can globally act on. Then a possible preprocessing operation is $\id_{A}\otimes\Lambda_{\rm sub}^{A'}[\rho^{AA'B}]=\rho^{AB}\otimes\ket{\psi}_{A'}\bra{\psi}$ (which can be realized acting on A' only).

Since the $\log d'_A$ contribution in \eqref{eq:capacity} can be considered purely classical, we choose a different way of counting the rate of transmission: indeed, it appears natural to subtract the logarithmic contribution in order to define the \emph{quantum advantage} of dense coding as
\beq
\label{eq:advantage}
\begin{split}
\Delta(A\>B)&\equiv S(\rho^B)-\inf_{\Lambda_A}S\big((\Lambda_A\otimes \id_B)[\rho_{AB}]\big)\\
						&=\sup_{\Lambda_A}I'(A\>B),
\end{split}
\eeq
where now it makes sense to consider the infimum (or the supremum) over all maps $\Lambda_A$, with whatever output dimension. Since a possible map is $\Lambda_{\rm sub}$, we have $\Delta(A\>B)\geq0$. We say that a state is Dense-Codeable (DC) if $\Delta(A\>B)$ is strictly positive. It may be that  $\Delta$ is not additive, hence to ensure that the state is not useful at all for dense coding, one has to consider its regularization (see Sec.~\ref{sec:manycopies}).

We remark that the classification of states in terms of their dense-codeability for different classes of encoding operations, will not depend on such redefinition. Moreover, the redefined quantity appears more information-theoretical and depends only on the state.

We now recall the analysis of a similar optimization problem which occurs in the study of entanglement of purification~\cite{IBMHor2002}. Since von Neumann entropy is concave, it is sufficient to consider extremal maps. The input of the map $\Lambda_A$ is an operator acting on a $d_A$-dimensional system, thus, according to~\cite{choi}, if $\Lambda_A$ is extremal it can be written by means of at most $d_A$ Kraus operators, i.e., as
\beq
\label{eq:map}
\Lambda_A[X]=\sum_{i=1}^{d_A}A_iXA_i^\dagger.
\eeq
The range of the operator $\Lambda[X]$ is given by all the columns of Kraus operators $A_i$, and each operator $A_i$ has $d_A$ (the input dimension) columns. Therefore, the optimal output dimension $d_{A'}$ can be taken to be $d_A^2$, and the infimum in \eqref{eq:advantage} is actually a minimum.  It is possible to further relate the quantum advantage of dense coding with entanglement of purification, as we do in Section \ref{sec:monogamy}.

Exploiting the convexity of entropy, it is immediate to find the following upper bound for $\Delta(A\>B)$:
\begin{gather}
\label{eq:Gbound}
\Delta(A\>B)\leq S(\rho^B)-\min_AS
\left(\rho^{AB}(A)
\right),\\
\nonumber\rho^{AB}(A)=\frac{A\otimes\openone\rho^{AB}A^\dagger\otimes\openone}{\Tr\big(A\otimes\openone\rho^{AB}A^\dagger\otimes\openone\big)}
\end{gather}
where, according to the reasoning of the previous paragraph, $A$ can be taken as a $d_A\times d_A$ square matrix. We remark that this is only an upper bound: local filtering is not allowed in our framework, because it requires classical communication. Moreover, with a true local filtering, the reduced density matrix $\rho^B$ changes, while we keep it fixed in \eqref{eq:Gbound}.

{\bf Example 1.} In \cite{bruss2} the dense-codeability by unitaries of the Werner state equivalent to
\beq
\label{eq:2qwerner}
\rho_p=p\pro{\psi_0}+(1-p)\frac{\openone}{4},
\eeq
with $\psi_0$ defined in \eqref{eq:maxent2qubits}, was studied. The state $\rho_p$ is entangled for $p>1/3$, but unitarily-DC only for $p>p_{\rm U-DC}=0.7476$. One may ask different questions: (i) Is the state $\rho_p$ DC for some $p<p_{\rm U-DC}$ if we allow general encoding operations, i.e., pre-processing? (ii) Can $\Delta(A\>B)$ be greater that $I(A\>B)$ when the latter is strictly positive? Question (i) addresses the problem of deciding whether a state that is not useful with some restricted encoding, is instead useful if we allow more general operations. Question (ii) addresses instead the problem of a ``greater usefulness'' by general encoding. In \cite{michal} numerical evidence was found that no pre-processing $\Lambda_A:\mathcal{M}(\mathbb{C}^2)\rightarrow\mathcal{M}(\mathbb{C}^2)$ can enhance $I(A\>B)$ in the case of a shared two-qubit state. However, note that, as previously discussed, optimal pre-processing maps are in principle of the form $\Lambda_A:\mathcal{M}(\mathbb{C}^2)\rightarrow\mathcal{M}(\mathbb{C}^4)$, i.e. with a larger output. Here we concentrate instead on the bound \eqref{eq:Gbound}, for which, as discussed, we can consider the matrix $A$ as $A:\mathbb{C}^2\rightarrow\mathbb{C}^2$. We observe that, thanks to the $U\otimes U^*$ symmetry of the state~\cite{reduction}, and to the invariance under unitaries of both the entropy and the trace, we can take $A$ to be diagonal in Alice's Schmidt basis for $\psi_0$, i.e. to be of the form $A=\begin{pmatrix} r & 0 \\ 0 & 1-r \end{pmatrix}$, with $0\leq r\leq 1$. It is possible to compute analytically the entropy
$S(\rho^{AB}(A))$. One finds that the optimal choice is $r=0,1$, and the bound is $1-H_2(\frac{1+p}{2})$, where $H_2(x)=-x \log x - (1-x)\log(1-x)$ is the binary entropy. Thus, in this example we see that the bound \eqref{eq:Gbound} is far from being tight, since it is strictly positive for every $p>0$, i.e. even for separable states. Therefore, it is not possible to use it to conclude something as regards question (i). Anyway, it constitutes a non-trivial limit on $\Delta(A\>B)$, and provides some information about question (ii).

\section{Preprocessing with many senders}
\label{sec:examples}

If the there are many senders, it may happen that the operations among them are restricted, for example no communication may be allowed, or they may collaborate only trough LOCC. Both latter situations do not affect the unitary encoding part of the dense coding. Indeed, it is possible to realize the optimal unitary operations locally: each Alice acts with unitaries satisfying the optimal condition on her subsystem~\cite{bruss,bruss2}.

We notice though that in the case of many senders and many receivers, even considering just unitary encoding, it happens that in certain cases local unitary encoding is not enough to take advantage of quantum correlations. This may be due, for example, to the fact that with restricted operations at the senders'  and at the receivers', the question of who sends what to whom---the network structure mentioned in Section~\ref{sec:introduction}---is really important. A very simple case where this is evident, is that of two senders $A$ and $A'$, and two receivers $B$ and $B'$, with $A$ and $B'$ ($A'$ and $B$) sharing an EPR pair, $A$ ($A'$) sending her quantum system to $B$ ($B'$), and both the senders and the receivers restricted to act by local operations. It is then clear that though the senders and the receivers share two EPR pairs, these are useless for the sake of dense coding, because they happen to pertain to the wrong pairs of senders and receivers. It is also obvious that the availability of global operations---or even just of the swap operation---on the senders' side would make dense coding possible, by using fully the quantum correlations existing between the set of senders and the set of receivers.

Considering the case of many senders and one receiver, but allowing encoding by general operations, one realizes that it might not be possible to apply the optimal pre-processing map. Indeed, just repeating the considerations which brought to \eqref{eq:capacity}, it is clear that, in the case of many senders and one receiver, the dense coding capacity may be expressed as
\beq
\chi^{d'_A}_O=\log d'_A + S(\rho^B)-\min_{\Lambda\in O}\big((\Lambda\otimes\id)[\rho^{AB}]\big),
\eeq 
where $O$ is the set of allowed operations on the senders' side, for example global (G)\footnote{It corresponds to the case where there is only one sender.}, LOCC, or LO. Obviously,
\beq
\label{eq:mshierarchy}
\chi^{d'_A}_{G}\geq \chi^{d'_A}_{LOCC}\geq \chi^{d'_A}_{LO}\geq \log d'_A.
\eeq
The capacity corresponds at least to the classical one with many senders and one receiver, because it is always possible for the Alices to apply locally the substitution map $\Lambda^A_{\rm sub}=\Lambda_{\rm sub}^{A_1}\otimes \cdots \otimes\Lambda_{\rm sub}^{A_N}$. The only subtle point is the compatibility of the choice of the target output dimension $d'_A$: we will suppose it is always of the factorized form $d'_A=d'_{A_1}\cdots d'_{A_N}$, so that it can be achieved exactly by an optimal local unitary encoding. Thus, we can define the corresponding (non-negative) quantum advantages
\beq
\Delta_G\geq\Delta_{LOCC}\geq\Delta_{LO}\geq 0.
\eeq

We can obtain an upper bound for $\Delta_{LOCC}$ -- and therefore valid also for $\Delta_{LO}$ -- similar to the presented in \eqref{eq:Gbound}:
\begin{multline}
\label{eq:LOCCbound}
%\begin{aligned}
\Delta_{LOCC}(A\>B)\\\leq S(\rho^B)
\,-\min_{A_{\rm prod}}S
\left(\frac{A_{\rm prod}\otimes\openone\rho^{AB}A_{\rm prod}^\dagger\otimes\openone}{\Tr\big(A_{\rm prod}\otimes\openone\rho^{AB}A_{\rm prod}^\dagger\otimes\openone\big)}
\right),
%\end{aligned}
\end{multline}
where $A_{\rm prod}=A_1\otimes\cdots\otimes A_N$, with each $A_i$ a $d_{A_i}\times d_{A_i}$ square matrix. 

\section{Examples of the hierarchy of capacities for multi-senders}
\label{sec:exampleshierarchy}

We provide examples of the hierarchy \eqref{eq:mshierarchy}, more precisely of shared states that are not DC for certain classes of allowed operations among senders, but are DC for more general operations. 
\subsection{LOCC-DC but not LO-DC}

We first analyze the case where the state is not LO-DC but it is LOCC-DC: $\Delta_{LO}=0$ while $\Delta_{LOCC}>0$. We will need the following lemma.

\begin{lemma}
Consider a tripartite state $\rho_{AA'B}$ 
such that (i) it is separable under the $A':AB$ cut, and (ii) its reduction $\rho_{AB}$ 
is also separable.  Then, after any bilocal operation $\Lambda_{AA'}=\Lambda_A\otimes\Lambda_{A'}$ of parties $A$ and $A'$, we have $I'(AA'\>B)\leq0$.
\end{lemma}
\begin{IEEEproof} For separable states coherent information 
is always non-positive \cite{HH94-redun}. For any state 
separable under $A':AB$ cut we then have 
\be
S(A'AB)\geq S(AB).
\ee
Now, if the state $\rho_{AB}$ is also separable, then 
\be
S(AB)\geq S(B).
\ee
Thus, for a tripartite state separable along the $A':AB$ cut and such that its $AB$ reduction is separable, $I(AA'\>B)\leq 0$.
Moreover, after any bilocal operation $\Lambda_A\otimes\Lambda_{A'}$ the state 
still satisfies the above separability features so that $I'(AA'\>B)\leq 0$.
\end{IEEEproof}

Note that the separability properties used in the previous lemma may not be preserved when the parties $AA'$ can communicate classically.

{\bf Example 2.}
Consider the state 
\be
\label{eq:exampleLOCCnonLO}
\rho_{AA'B}= \frac{1}{2} (|\phi_0 \>\<\phi_0|+|\phi_1 \>\<\phi_1|),
\ee
where 
\begin{align}
\ket{\phi_0}&= |0\>_{A'}\otimes \frac{1}{\sqrt 2} (|00\>_{AB} + |11\>_{AB})\\
\ket{\phi_1}&= |1\>_{A'}\otimes \frac{1}{\sqrt 2} (|01\>_{AB} + |10\>_{AB}),
\end{align}
which has initial entropies $S(AA'B)=S(B)=1$ and $I(AA'\>B)=0$.
The state is explicitly separable with respect to the $A':AB$ cut. The trace over $A'$ gives an equal mixture 
of two qubit orthogonal maximally entangled states, hence it is separable~\cite{BDSW1996}. 
Thus, according to Lemma 1, parties $A$ and $A'$ 
cannot locally decrease the total entropy below $S(B)$. However by LOCC they can. Namely, $A'$ 
can measure in the $\{0,1\}$ basis, communicate the result to $A$, and further substitute the subsystem $A'$ with one in a pure state. Then, after a suitable local unitary rotation, $A$ will share a maximally entangled state with $B$ and $S'(AA'B)=0$, so that $I'(AA'\>B)=1$.

The previous example is the simplest possible one that illustrates how some ``spread'' noise which conflicts with (unitary) dense coding can be undone only allowing operations among the senders that are more general than local operations. Indeed, in the system $AA'$ one can single out a \emph{virtual} qubit, carrying the whole noise. The noisy qubit is however encoded non-locally into the system $AA'$, so that the senders do not have \emph{local} access to it. Effective tracing out of the unwanted noise (prior to unitary encoding) is possible only if $A$ and $A'$ communicate. Indeed, one can go from $\rho_{AA'B}$ to $\rho_{\tilde{A}\tilde{A}'B}=\openone_{\tilde{A}'}/2\otimes\psi_0^{\tilde{A}B}$ by an invertible $A:A'$-LOCC operation, but not by an $A:A'$-LO operation. 

The previous tripartite (two senders, one receiver) case can be generalized straightforwardly. Following the definition of multipartite mutual information, one can define a quantity,
which is not an entanglement measure, but may be useful (see~\cite{entpar} in this context):
\beq
\begin{aligned}
D(B:A_1: \ldots :A_n)&\equiv E(B:A_1) + E(BA_1:A_2) \\
                                     &\quad+\ldots\\
                                     &\quad+E(BA_1A_2\ldots A_{n-1}:A_n),
\end{aligned}
\ee
where $E$ is any entanglement parameter, i.e. it is positive and $E(X:Y)=0$ if and only if the state $\rho_{X:Y}$ is separable. Similarly as in tripartite case, we obtain that if $D$ is zero, then parties 
$A_1\ldots A_{n}$ cannot make the global entropy be less than $S(B)$ by LO (but not necessarily by LOCC), so that the state is useless for superdense coding from $A_1\ldots A_n$ to B.

{\bf Example 3.}
Consider the state
\begin{multline}
\rho_{A_1\ldots A_nB}\\
\begin{aligned}
&=\frac{1}{2^{n-1}}\sum_{i_2,\ldots,i_n=0}^{1}\pro{i_2}\otimes\cdots\otimes\pro{i_n}\\
&\qquad\qquad\otimes
\Big(\big(\sigma_{\bigoplus_{j=2}^ni_j}^{A_1}\otimes\openone^B\big)P^0_{A_1B}\big(\sigma_{\bigoplus_{j=2}^ni_j}^{A_1}\otimes\openone^B\big)\Big),
\end{aligned}
\end{multline}
where $\sigma_0$ and $\sigma_1$ are the identity and the flip operator, respectively, $P^0$ is the projector onto the maximally entangled state $\ket{\psi_0}$, and $\oplus$ corresponds to addition modulo 2. It easily checked that $\rho_{A_1\ldots A_nB}$ satisfies $D=0$. The unitary rotation $\sigma_a$ applied to the $A_1$ part of the maximally entangled state depends on the ``parity'' of the state of the other Alices. It is correctly identified if all A$_2$, \ldots, A$_n$ measure their qubits in the computational basis, and communicate their results to $A_1$, which can then share a singlet with $B$.

\subsection{G-DC but not LOCC-DC}

To have an example of a state for which $\Delta_{LOCC}=0$ while $\Delta_{G}>0$, consider the Smolin state \cite{Smolin-unlock} 
\be
\label{eq:smolin}
\rho_{A_1BA_2A_3}=
\sum_{\mu=0}^3|\psi_\mu\>_{A_1B}\<\psi_\mu| \ot |\psi_\mu\>_{A_2A_3}\<\psi_\mu|
\ee
where $\psi_\mu$ are Bell states.
Note that states $\psi_\mu$ are indistinguishable by LOCC \cite{BellPRL}. 
Hence it seems reasonable, that the state cannot be used for super dense coding, 
even if the parties $A_1$, $A_2$ and $A_3$ can use LOCC. For example, the parties $A_2A_3$ 
cannot distinguish which Bell state they have, hence cannot tell $A_1$ 
what rotation to apply, in order to share singlet with $B$. Let us now prove 
that this is true.  

The state is $A_1A_2:BA_3$ separable (from Eq. \eqref{eq:smolin} it is explicitly $A_1B:A_2A_3$ separable, 
however it is permutationally invariant~\cite{Smolin-unlock}). After any LOCC operation 
this will not change. Thus the output state of systems $A_1B$ 
will be separable, hence $S'(A_1B)\geq S'(B)=S(B)$. Moreover the 
total output state will remain $A_1B:A_2A_3$ separable, which implies 
$S'(A_1BA_2A_3) \geq S'(A_1B)$. Combining the two inequalities we get 
\be
S'(A_1BA_2A_3) \geq S(B).
\ee
Of course, if for example $A_2$ and $A_3$ could meet and perform global operations, the state would become
useful for dense coding, as they could help $A_1$ to share a singlet with $B$. 

\section{Limits on pre-processing from one-way distillability and symmetric extensions}
\label{sec:symmext}

The possibility of global pre-processing makes  non-trivial the identification of states which, although $A_1\ldots A_n:B$ is entangled, are not G-DC ($\Delta_G\leq0$). One has to exclude that the coherent information can be made strictly positive by any action on the side of Alices. We will now see how this may be related to one-way distillation and the concept of symmetric extension. Since we will focus on global operations, we may as well consider a bipartite setting.

Loosely speaking, entanglement distillation consists of the process of obtaining $m$ copies of the highly entangled pure states \eqref{eq:maxent2qubits}, starting from $n$ copies of a mixed entangled state, by means of a restricted class of operations that can not create entanglement~\cite{BDSW1996}. The optimal rate, i.e. the optimal ratio $m/n$, of the conversion for $n$ that goes to infinity, is the \emph{distillable entanglement} under the given constraint on operations. The class may be chosen to be LOCC operations -- in such case we speak simply of distillable entanglement -- or, more restrictively, one-way LOCC operations, for which classical communication is allowed only from one party to the other, and not in both directions. In the latter case we speak of one-way distillable entanglement. If we suppose that the communication goes from Alice to Bob, it has been showed~\cite{hashinginequality} that the one-way distillable entanglement $E_D(A\>B)$ of a state $\rho^{AB}$ satisfies the hashing inequality
\beq
E_D(A\>B)\geq I(A\>B),
\eeq
hence
\beq
E_D(A\>B)\geq \Delta(A\>B),
\eeq
i.e., it is greater than the quantum advantage of dense coding. It follows that any DC state is not only distillable, but even one-way distillable. In turn, if a state is not one-way distillable, then it can not be DC.

There are entangled states for which we know $E_D(A\>B)=0$: states which admit $B$-symmetric extensions~\cite{DohertyetalPRL,DohertyetalPRA,terhalPRLext}. A state $\rho^{AB}$ admits a $B$-symmetric extension if there exists a state $\sigma^{ABB'}$ such that its reductions satisfy
\[
\sigma^{AB}=\sigma^{AB'}=\rho^{AB}.
\]
Suppose $\rho^{AB}$ has a tripartite symmetric extension $\sigma^{ABB'}$ and is at the same time one-way distillable. A one-way distillation protocol consists of an Alice operation whose result -- the index of the Kraus operator in \eqref{eq:map} -- is communicated to the other party.  Bob can then perform an operation depending on the result received; no further action of Alice is required. The communication involved is classical, so it can be freely sent to many parties. If, having at disposal  $\sigma^{ABB'}$, we run the one-way LOCC protocol, which by hypothesis allows distillation, in parallel between $A$ and $B$, and $A$ and $B'$, we would end up with a subsystem $A$ which is at the same time maximally entangled both with $B$ and $B'$. However, this is impossible, because of monogamy of entanglement~\cite{CKW}. We conclude that a one-way distillable state does not admit a symmetric $B$-extension~\cite{PHsymmetric}.
As regards the case of the two-qubit Werner state~\eqref{eq:2qwerner}, in~\cite{acinsymm}, it was proved that it admits a symmetric extension for $p\leq 2/3$.

\section{Limits on many-copies processing}
\label{sec:manycopies}

The examples of the classification we discussed in Section~\ref{sec:exampleshierarchy} depend only on relations among entropies which rely on separability properties. As such, the action on many copies of the state at disposal can not help. Indeed, following~\cite{michal,winterdense}, one can define the quantum advantage per copy when the encoding is allowed on $n$-copies of the state at the same time:
\beq
\begin{split}
\Delta^{(n)}(A\>B)&=\frac{1}{n}\Delta(A\>B)_{\rho_{AB}^{\otimes n}}\\
									&=S(\rho^B)-\frac{1}{n}\min_{\Lambda_A^{(n)}}
									S\big((\Lambda^{(n)}_A\otimes \id_B)[\rho_{AB}^{\otimes n}]\big),
\end{split}
\eeq
where now $\Lambda_A^{(n)}$ acts on $\mathcal{M}(\mathbb{C}^{d_A^n})$, and the asymptotic quantum advantage per copy:
\beq
\label{eq:asymp_advantage}
\begin{split}
\Delta^\infty(A\>B)&=\lim_{n\rightarrow\infty}\Delta^{(n)}(A\>B).
\end{split}
\eeq
Correspondingly, one has the multipartite quantum advantages $\Delta^{(n)}_O(A\>B)$ and $\Delta^{\infty}_O(A\>B)$ where the Alices are restricted to the class of operations $O$. It is clear that
$\Delta^{(n)}_{LO}(A\>B)=\Delta^{\infty}_{LO}(A\>B)=0$ and $0<\Delta^{(n)}_{LOCC}(A\>B)\leq\Delta^{\infty}_{LOCC}(A\>B)$ for the state~\eqref{eq:exampleLOCCnonLO}, while $\Delta^{(n)}_{LOCC}(A\>B)=\Delta^{\infty}_{LOCC}(A\>B)=0$ and $0<\Delta^{(n)}_{G}(A\>B)\leq\Delta^{\infty}_{G}(A\>B)$ for the state~\eqref{eq:exampleLOCCnonLO} for the Smolin state \eqref{eq:smolin}.

\section{Monogamy relation between entanglement of purification and the advantage of dense coding}
\label{sec:monogamy}

We observed in Section \ref{sec:two-party} that there are similarities in the calculation of the advantage of dense coding and in that of entanglement of purification \cite{IBMHor2002}. In this section we will see that this relation is more than a coincidence: there is in fact a monogamy relation between the advantage of dense coding and the entanglement of purification, that does not seem to have already been reported in  literature.

We start by recalling the definition of entanglement of purification for a bipartite state $\rho_{AB}\in\mathcal{M}(\mathbb{C}^{d_A})\otimes\mathcal{M}(\mathbb{C}^{d_B})$:
\beq
E_p(\rho_{AB})=E_p(A:B)=\min_{\psi:\tr_{A'B'}(\psi)=\rho_{AB}} S(\psi_{AA'}),
\eeq
where the minimum runs over all purifications $\psi=\pro{\psi}_{AA'BB'}$, $\ket{\psi}\in\mathbb{C}^{d_A}\otimes\mathbb{C}^{d_{A'}}\otimes\mathbb{C}^{d_B}\otimes\mathbb{C}^{d_{B'}}$ such that $\tr_{A'B'}(\psi)=\rho_{AB}$. Entanglement of purification is a measure of total correlations, where all correlations---even those of separable states---are somehow thought as being due to entanglement. Indeed, in the bipartite pure-state case, the entropy of one subsystem is an entanglement measure~\cite{BBPS1996}. For $\{\lambda_i,\pro{\lambda_i}\}$ the spectral ensemble of $\rho_{AB}$, consider its purification $\ket{\tilde\psi}=\sum_i\sqrt{\lambda_i} \ket{\lambda_i}_{AB}\ket{i}_{A'}\ket{0}_{B'}=\ket{\psi}_{AA'B}\ket{0}_{B'}$. Then any other purification can be obtained from $\ket{\tilde{\psi}}$ by means of an isometry $U_{A'B'}$ as $\ket{\psi}=U_{AB}\otimes\openone_{A'B'}\ket{\tilde{\psi}}$. Following \cite{IBMHor2002}, we find
\begin{align*}
\psi_{AA'}&=\tr_{BB'}(\psi_{AA'BB'})\\
	&=\tr_{BB'}(U_{A'B'}\tilde\psi_{AA'B}\otimes\pro{0}_{B'}U_{A'B'}^\dagger)\\
	&=(\Lambda_{A'}\otimes\id_{A})[\tr_B(\tilde\psi_{AA'B})]\\
	&=(\Lambda_{A'}\otimes\id_{A})[\tilde\psi_{AA'}]
\end{align*}
where $\Lambda_{A'}[X_{A'}]=\tr_{B'}(U_{A'B'}X_{A'}\otimes\pro{0}_{B'}U_{A'B'}^\dagger)$, for all $X_{A'}\in\mathcal{C}(\mathbb{C}^{d_{A'}})$. By varying $U_{A'B'}$, that is the purification $\psi$, we vary $\Lambda_{A'}$.
Thus,
\beq
\label{eq:entpurmap}
E_p(\rho_{AB})\equiv\min_{\Lambda_{A'}}S\big((\Lambda_{A'}\otimes \id_A)[\tilde\psi_{AA'}]\big),
\eeq
Comparing \eqref{eq:advantage} and \eqref{eq:entpurmap} we then conclude that, given a pure tripartite state $\psi_{ABC}$, one has
\beq
\label{eq:monogamypur}
S(B)=\Delta(A\rangle B)+E_p({B:C}).
\eeq
For fixed entropy $S(B)$, this means that the more $B$ is correlated with $C$, the less dense coding is advantageous from $A$ to $B$.
For a tripartite \emph{mixed} state $\rho_{ABC}$, following \cite{koashi2004monogamy} we may consider a purification $\psi_{ABCD}$, and apply equation \eqref{eq:monogamypur} to the three parties $(AD)$, $B$ and $C$ to find
\[
S(B)\geq\Delta(A\rangle B)+E_p({B:C}).
\]
Indeed, from the definition of advantage it is easy to check that $\Delta(AD\rangle B)\geq \Delta(A\rangle B)$ for all tripartite states $\rho_{ABD}$, in particular for the $ABD$ reduction of $\psi_{ABCD}$.
Following \cite{koashi2004monogamy} again, we may consider the asymptotic case, applying the just found relations to $\psi_{ABC}^{\otimes n}$ ($\rho_{ABC}^{\otimes n}$), using the additivity of von Neumann entropy, dividing by $n$, and taking the limit $n\rightarrow\infty$, to find
\beq
S(B)=\Delta^{\infty}(A\rangle B)+E_{LOq}({B:C}),
\eeq
and
\beq
S(B)\geq\Delta^{\infty}(A\rangle B)+E_{LOq}({B:C}),
\eeq
for the case of pure and mixed states, respectively. Here $E_{LOq}({A:B})=\lim_{n}\frac{1}{n}E_p(\rho_{AB}^{\otimes n})$ is the cost---in singlets---to create $\rho_{AB}$ in
the asymptotic regime, allowing approximation, from an initial supply of EPR-pairs by means of local operations and
asymptotically vanishing communication~\cite{IBMHor2002}.

For the pure state case it is fascinating to put together the results of Theorem 1 of \cite{koashi2004monogamy} and the present ones, to find relations between different notions of correlations and entanglement measures/parameters:
\begin{align}
\label{eq:monogamysingle}I_{\rm HV}(A\rangle B)-\Delta(A\rangle B)&=E_p({B:C})-E_F(B:C),\\
\label{eq:monogamymany}C_D(A\rangle B)-\Delta^\infty(A\rangle B)&=E_{LOq}({B:C})-E_C(B:C),
\end{align}
where, for a bipartite state $\rho_{AB}$:
\begin{itemize}
\item $I_{HV}$ is the measure of correlations defined in \cite{HendersonVedral} as
\[
I_{\rm HV}(A\rangle B)=\max_{\{M_x\}}\big[S(\rho_B)-\sum_xp_xS(\rho_B^x)\big],
\]
where the maximum is taken over all the POVMs $\{M_x\}$ applied on system $A$, $p_x\equiv\tr((M_x\otimes\openone)\rho_{AB})$ is the probability of the outcome $x$, $\rho_B^x\equiv\tr_A((M_x\otimes\openone)\rho_{AB}) /p_x$ is the conditional state on $B$ given the outcome $x$ on $A$, and $\rho_B=\sum_xp_x\rho_B^x=\tr_A(\rho_{AB})$;
\item
$C_D$ is the common randomness distillable by means of one-way classical communication from $A$ to $B$, that is the net amount of correlated classical bits that $A$ and $B$ can asymptotically share starting from an initial supply of copies of $\rho_{AB}$; it is equal to $C_D(A\rangle B)=\lim_{n}\frac{1}{n}I_{\rm HV}(A\rangle B)_{\rho_{AB}^{\otimes n}}$~\cite{DevetakW03-common};
\item $E_F$ is the entanglement of formation
\[
E_F(A:B)=\min_{\{(p_i,\psi^{AB}_i)\}} \sum_ip_iS(\psi_i^A),
\]
where the minimum runs over all pure ensembles such that $\sum_ip_i\psi_i^{AB}=\rho_{AB}$
\item $E_C$ is the entanglement cost,  that is the cost---in singlets---to create $\rho_{BC}$ in
the asymptotic regime, allowing approximation, from an initial supply of EPR-pairs by means of local operations and classical communication; it is equal to $E_C(A:B)=\lim_{n}\frac{1}{n}E_F(A:B)_{\rho_{AB}^{\otimes n}}$~\cite{cost}.
\end{itemize}

Note that the differences appearing in \eqref{eq:monogamysingle} and \eqref{eq:monogamymany} are positive~\cite{IBMHor2002}.

\section{Discussion}
\label{sec:discussion}

The difference with the results presented in~\cite{michal,winterdense} is two-fold. Firstly, in defining the quantum advantage of dense coding, we immediately consider a maximum over maps without restricting the dimension of the output. This means that we focus on the property of the state, rather than of a couple state+channel. Secondly, exactly for the same reason, we do not distinguish between many uses of the state and many uses of the channel: the rate is always defined in terms of the number of copies of the state used, even when we allow encoding on many copies. This two facts make our quantities $\Delta$ and $\Delta^\infty$ different from all the ones presented in~\cite{michal,winterdense}. In particular, we claim that the quantity $\Delta^\infty$ is more information theoretical than the quantity
\beq
\label{eq:genenc}
\overline{DC^{(\infty)}}(\rho)=1+\sup_n\sup_{\Lambda_A}\frac{nS(\rho^B)-S\big((\Lambda_A\otimes\id^{\otimes n})[\rho^{\otimes n}]\big)}{S\big(\rho_A^{\otimes n}\big)},
\eeq
which, according to~\cite{michal,winterdense}, corresponds to the rate of classical communication per qubit sent, i.e. per use of a two-dimensional quantum channel. Indeed, in the latter case one considers the use of whatever number of copies of the shared state per use of the channel. In particular, we remark that for pure states one has $\overline{DC^{(\infty)}}(\psi^{AB})=2$ as soon as the state $\psi^{AB}$ is entangled---whatever the degree of its entanglement---while $\Delta(\psi^{AB})=S(\rho^B)$.

Further, we notice that the distinction of usefulness of states for dense coding according to the allowed encoding operations, holds also for the quantities presented in~\cite{michal,winterdense}, as it is evident, for example, from~\eqref{eq:genenc}.

In conclusion, we considered the transmission of classical information by exploiting (many copies of) a shared quantum state, both in the bipartite and in the multipartite -- more specifically, in the many-to-one -- setting.
We discussed fundamental limits on the usefulness of states for multipartite dense coding, for given constraints on the operations allowed among senders. Such limits are not removed even if we allow the most general encoding under whatever number of copies of the shared state. Such analysis leads to a non-trivial classification of quantum states, parallel to the one suggested in~\cite{bruss, bruss2}, where constraints on the operations allowed on the receivers side (in a many-to-many communication setting) were considered. Indeed, one can depict a subdivision of multipartite states into classes of states that are many-to-one dense-codeable if certain operations, for example LOCC, are allowed among the senders, but not if the senders are restricted to local operations. 

Finally, focussing on general properties of dense-codeability of states, we observed that there exist a monogamy relation between the quantum advantage of dense coding and the entanglement of purification. Such a relation puts in quantitative terms the fact that the quantum advantage of dense coding is (or can be) large (only) if the disorder---as quantified by the von Neumann entropy---of the receiver is due to correlations with the sender, rather than with a third party. 

\section*{Acknowledgement}
We thank D. Bruss, K. Horodecki, P. Horodecki, R. Horodecki, C. Mora, A. Sen(De), U. Sen and J. Oppenheim for useful discussions. We gratefully acknowledge funding from EU (RESQ (IST 2001 37559), IP SCALA). M. P. acknowledges support from CNR-NATO.

%bibliographystyle{IEEEtran}
%\bibliography{IEEEabrv,biblio_sep-2009,dense2009}

\begin{thebibliography}{10}
\providecommand{\url}[1]{#1}
\csname url@samestyle\endcsname
\providecommand{\newblock}{\relax}
\providecommand{\bibinfo}[2]{#2}
\providecommand{\BIBentrySTDinterwordspacing}{\spaceskip=0pt\relax}
\providecommand{\BIBentryALTinterwordstretchfactor}{4}
\providecommand{\BIBentryALTinterwordspacing}{\spaceskip=\fontdimen2\font plus
\BIBentryALTinterwordstretchfactor\fontdimen3\font minus
  \fontdimen4\font\relax}
\providecommand{\BIBforeignlanguage}[2]{{%
\expandafter\ifx\csname l@#1\endcsname\relax
\typeout{** WARNING: IEEEtran.bst: No hyphenation pattern has been}%
\typeout{** loaded for the language `#1'. Using the pattern for}%
\typeout{** the default language instead.}%
\else
\language=\csname l@#1\endcsname
\fi
#2}}
\providecommand{\BIBdecl}{\relax}
\BIBdecl

\bibitem{horodecki2009quantum}
R.~Horodecki, P.~Horodecki, M.~Horodecki, and K.~Horodecki, ``Quantum
  entanglement,'' \emph{Reviews of Modern Physics}, vol.~81, no.~2, pp.
  865--942, 2009.

\bibitem{NC}
M.~A. Nielsen and I.~L. Chuang, \emph{Quantum Computation and Quantum
  Information}.\hskip 1em plus 0.5em minus 0.4em\relax Cambridge: Cambridge
  University Press, 2000.

\bibitem{Alber2001}
G.~Alber, T.~Beth, M.~Horodecki, P.~Horodecki, R.~Horodecki, M.~Rotteler,
  H.~Weinfurter, R.~Werner, and A.~Zeilinger, \emph{Quantum Information: An
  Introduction to Basic Theoretical Concepts and Experiments}.\hskip 1em plus
  0.5em minus 0.4em\relax Springer, 2001.

\bibitem{teleportation}
C.~H. Bennett, G.~Brassard, C.~Crepeau, R.~Jozsa, A.~Peres, and W.~K. Wootters,
  ``Teleporting an unknown quantum state via dual classical and
  einstein-podolsky-rosen channels,'' \emph{Phys. Rev. Lett}, vol.~70, p. 1895,
  1983.

\bibitem{BennettWiesner}
C.~Bennett and S.~Wiesner, ``Communication via one- and two-particle operators
  on einstein-podolsky-rosen states,'' \emph{Phys. Rev. Lett}, vol.~69, p.
  2881, 1992.

\bibitem{holevo}
A.~S. {Holevo (Kholevo)}, ``Bounds for the quantity of information transmitted
  by a quantum communication channel,'' \emph{Probl. Peredachi Inf.}, vol.~9,
  p.~3, 1973, translated in Problems Inf. Transmiss. 9, 177Ð183 (1973).

\bibitem{BBPS1996}
C.~H. Bennett, H.~Bernstein, S.~Popescu, and B.~Schumacher, ``Concentrating
  partial entanglement by local operations,'' \emph{Phys. Rev. A}, vol.~53, pp.
  2046--2052, 1996.

\bibitem{bruss}
D.~Bru\ss, G.~D'Ariano, M.~Lewenstein, C.~Macchiavello, A.~Sen(De), and U.~Sen,
  ``{Distributed quantum dense coding},'' \emph{Phys. Rev. Lett}, vol.~93, p.
  210501, 2005.

\bibitem{bruss2}
D.~Bru{\ss}, G.~D'Ariano, M.~Lewenstein, C.~Macchiavello, A.~Sen(De), and
  U.~Sen, ``Dense coding with multipartite quantum states,'' \emph{Int. J.
  Quant. Inf.}, vol.~4, p. 415, 2006.

\bibitem{SchumNiel}
B.~Schumacher and M.~A. Nielsen, ``Quantum data processing and error
  correction,'' \emph{Phys. Rev. A}, vol.~54, p. 2629, 1996.

\bibitem{HH94-redun}
R.~Horodecki and P.~Horodecki, ``Quantum redundancies and local realism,''
  \emph{Phys. Lett. A}, vol. 194, pp. 147--152, 1994.

\bibitem{hashinginequality}
I.~Devetak and A.~Winter, ``Distillation of secret key and entanglement from
  quantum states,'' \emph{Proc. R. Soc. Lond. A}, vol. 461, p. 207, 2005.

\bibitem{michal}
M.~Horodecki, P.~Horodecki, R.~Horodecki, D.~W. Leung, and B.~Terhal,
  ``Classical capacity of a noiseless quantum channel assisted by noisy
  entanglement,'' \emph{Quantum Inf. Comput.}, vol.~1, p.~70, 2001.

\bibitem{winterdense}
A.~Winter, ``Scalable programmable quantum gates and a new aspect of the
  additivity problem for the classical capacity of quantum channels,'' \emph{J.
  Math. Phys.}, vol.~43, p. 4341, 2002.

\bibitem{IBMHor2002}
B.~M. Terhal, M.~Horodecki, D.~P. DiVincenzo, and D.~Leung, ``The entanglement
  of purification,'' \emph{J. Math. Phys.}, vol.~43, pp. 4286--4298, 2002.

\bibitem{choi}
M.~Choi, ``Completely positive linear maps on complex matrices,'' \emph{Linear
  Algebra Appl.}, vol.~10, p. 285, 1975.

\bibitem{reduction}
M.~Horodecki and P.~Horodecki, ``Reduction criterion for separability and
  limits for a class of protocol of entanglement distillation,'' \emph{Phys.
  Rev. A}, vol.~59, pp. 4206--4216, 1999.

\bibitem{BDSW1996}
C.~H. Bennett, D.~P. DiVincenzo, J.~Smolin, and W.~K. Wootters, ``Mixed-state
  entanglement and quantum error correction,'' \emph{Phys. Rev. A}, vol.~54,
  pp. 3824--3851, 1996.

\bibitem{entpar}
B.~Synak-Radtke, {\L}.~Pankowski, M.~Horodecki, and R.~Horodecki, ``On some
  entropic entanglement parameter,'' e-print quant-ph/0608201.

\bibitem{Smolin-unlock}
J.~A. Smolin, ``Four-party unlockable bound entangled state,'' \emph{Phys. Rev.
  A}, vol.~63, p. 032306, 2001.

\bibitem{BellPRL}
S.~Ghosh, G.~Kar, A.~Roy, A.~Sen(De), and U.~Sen, ``Distinguishability of bell
  states,'' \emph{Phys. Rev. Lett}, vol.~87, p. 277902, 2001.

\bibitem{DohertyetalPRL}
A.~C. Doherty, P.~A. Parillo, and F.~M. Spedalieri, ``Distinguishing separable
  and entangled states,'' \emph{Phys. Rev. Lett.}, vol.~88, p. 187904, 2002.

\bibitem{DohertyetalPRA}
------, ``Complete family of separability criteria,'' \emph{Phys. Rev. A},
  vol.~69, p. 022308, 2004.

\bibitem{terhalPRLext}
A.~C.~D. B.~M.~Terhal and D.~Schwab, ``Symmetric extensions of quantum states
  and local hidden variable theories,'' \emph{Phys. Rev. Lett.}, vol.~90, 2003.

\bibitem{CKW}
V.~Coffman, J.~Kundu, and W.~K. Wootters, ``Distributed entanglement,''
  \emph{Phys. Rev. A}, vol.~61, p. 052306, 2000.

\bibitem{PHsymmetric}
P.~Horodecki and M.~{\L}. Nowakowski, ``A simple test for quantum channel
  capacity,'' \emph{J. Phys. A: Math. Theor.}, vol.~42, p. 135306, 2009.

\bibitem{acinsymm}
G.~T{\'o}th and A.~Ac{\'i}n, ``Genuine tripartite entangled states with a local
  hidden-variable model,'' \emph{Phys. Rev. A}, vol.~74, p. 030306, 2006.

\bibitem{koashi2004monogamy}
M.~Koashi and A.~Winter, ``{Monogamy of quantum entanglement and other
  correlations},'' \emph{Physical Review A}, vol.~69, no.~2, p. 22309, 2004.

\bibitem{HendersonVedral}
L.~Henderson and V.~Vedral, ``Classical, quantum and total correlations,''
  \emph{J. Phys. A}, vol.~34, pp. 6899--6905, 2001.

\bibitem{DevetakW03-common}
I.~Devetak and A.~Winter, ``Distilling common randomness from bipartite quantum
  states,'' \emph{IEEE Trans. Inf. Theory}, vol.~50, p. 3183, 2004.

\bibitem{cost}
P.~Hayden, M.~Horodecki, and B.~Terhal, ``The asymptotic entanglement cost of
  preparing a quantum state,'' \emph{J. Phys. A}, vol.~34, pp. 6891--6898,
  2001.

\end{thebibliography}

% Generated by IEEEtran.bst, version: 1.13 (2008/09/30)

\end{document}